\documentclass[aps,prd,twocolumn,superscriptaddress,nofootinbib]{revtex4}

\usepackage{colordvi} 
\usepackage{bm}
\usepackage{multirow}

\usepackage{amsmath,latexsym}    
\usepackage{graphicx}   
\usepackage{verbatim}   
\usepackage{xcolor}      
\usepackage{subfigure}  
\usepackage{hyperref}   
\usepackage{tensor}     
\usepackage{amssymb}
\usepackage{enumerate}	
\usepackage{braket}		

\newcommand{\hompc}{\, h\, {\rm Mpc}^{-1}}
\newcommand{\mpcoh}{\,h^{-1}\,{\rm Mpc}}
\newcommand{\moh}{\,h^{-1}\,M_{\odot}}
\newcommand{\Gpcoh}{\,h^{-1}\,{\rm Gpc}}

\newcommand{\be}{\begin{equation}}
\newcommand{\ee}{\end{equation}}
\newcommand{\bs}{\rm bs}
\newcommand{\real}{\rm real}

\newcommand{\fs}{f\sigma_{8}}
\newcommand{\fzsz}{f(z)\sigma_{8}(z)}

\newcommand{\bfk}{\boldsymbol{k}}
\newcommand{\bfr}{\boldsymbol{r}}
\newcommand{\bfx}{\boldsymbol{x}}

\newcommand{\bfv}{\boldsymbol{v}}

\newcommand{\kmin}{k_{\rm min}}
\newcommand{\rockstar}{\texttt{ROCKSTAR} }
\newcommand{\darkquest}{\texttt{DARK QUEST} }

\newcommand{\xivv}{\xi_{\rm{v}}}

\newcommand{\Pmm}{P_{\rm{m}}}

\newcommand{\Cext}{\mathbf{C}_{\rm ext}}
\newcommand{\Cint}{\mathbf{C}_{\rm int}}

\newcommand{\LoneG}{L_{1\rm G}}
\newcommand{\LtwoG}{L_{2\rm G}}
\newcommand{\LoneGsub}{L_{2{\rm G}}^{\times 0.5}}
\newcommand{\Lthsub}{L_{2{\rm G}}^{\times 0.15}}

\newcommand{\titlename}{
Large-scale halo velocity correlations and the impact of finite simulation volumes}

\newcommand{\ntu}{Department of Physics, National Taiwan University (NTU), No. 1, Section 4, Roosevelt Road, Taipei 10617, Taiwan}
\newcommand{\asiaa}{Academia Sinica Institute of Astronomy and Astrophysics (ASIAA), No. 1, Section 4, Roosevelt Road, Taipei 106319, Taiwan}
\newcommand{\ipmu}{Kavli IPMU (WPI), UTIAS, The University of Tokyo, Kashiwa, Chiba 277-8583, Japan}
\newcommand{\yitp}{Center for Gravitational Physics and Quantum Information, Yukawa Institute for Theoretical Physics, Kyoto University, Kyoto 606-8502, Japan}

\newcommand{\kyosan}{Department of Astrophysics and Atmospheric Sciences, Faculty of Science, Kyoto Sangyo University, Kyoto 603-8555, Japan}
\newcommand{\riken}{RIKEN Center for Advanced Intelligence Project, 1-4-1 Nihonbashi, Chuo-ku, Tokyo 103-0027, Japan}

\begin{document}

\title{
\titlename
}

\author{Yao-Tsung Chuang}
\affiliation{\asiaa}
\affiliation{\ntu}

\author{Teppei Okumura}\email{tokumura@asiaa.sinica.edu.tw}
\affiliation{\asiaa}
\affiliation{\ipmu}

\author{Takahiro Nishimichi}
\affiliation{\kyosan}
\affiliation{\riken}
\affiliation{\yitp}
\affiliation{\ipmu}

\date{\today}


\begin{abstract}
The velocity correlation functions directly measured from the peculiar velocity field of dark matter in numerical simulations are known to have an amplitude lower than that predicted by theoretical models at large scales. The trend persists for dark-matter halos or galaxies that are more closely related to the observables. We investigate the impact of the finite simulation box sizes on the measured velocity correlation functions of halos, utilizing $N$-body simulations with different box sizes. We measure the halo velocity correlations from $N$-body simulations with side lengths of $1\,\Gpcoh$ and $2\,\Gpcoh$, confirming the former is more suppressed compared to the linear theory prediction on large scales due to the lack of large-scale modes beyond the box size. In contrast, even though we subdivide the larger-box simulations into those with side lengths of $1\Gpcoh$, the amount of the suppression is the same as that from the original boxes, as the large-scale modes are already imprinted. Introducing the lower limit of the integral in the Hankel transform, $\kmin$, as a free parameter and marginalizing it over, we find that the constrained growth rate parameter, $\fzsz$, returns the correct value assumed in the simulations. However, when we ignore the effect and set $\kmin=0$, the constraint on $\fs$ is significantly biased if the correlation between different separation bins is also ignored. Furthermore, we find that the suppression of the velocity correlation amplitude on large scales depends on halo mass, with more massive halos exhibiting a systematically stronger suppression. These results highlight the importance of accounting for missing long-wavelength modes when developing simulation-based modeling of velocity statistics, such as emulators. 
\end{abstract}

\maketitle
{\it Introduction.}
The large-scale structure of the Universe emerged from the gravitational growth of primordial density fluctuations \cite{Peebles:1980}. While galaxy clustering has long been the primary observable used to study this process, the peculiar velocity field offers a complementary and more direct probe of gravitational dynamics, as velocities respond to the total matter distribution rather than its biased tracers \cite{Strauss:1995}.

In galaxy redshift surveys, peculiar velocities are encoded in the large-scale structure as anisotropies in galaxy clustering, known as redshift-space distortions (RSD), and have been widely used to constrain the growth rate of the universe \cite{Kaiser:1987,Hamilton:1992,Scoccimarro:2004,Guzzo:2008,Okumura:2011}. Beyond RSD, galaxy and cluster peculiar velocities can also be probed directly through distance-redshift comparisons \cite{Bhattacharya:2008,Feldman:2010}. The spatial correlations of these velocities are particularly sensitive to large-scale modes and provide a powerful probe of cosmology, enabling precise constraints on the growth rate, $\fzsz$, and tests of gravity on cosmological scales \cite{Koda:2014,Okumura:2014,Turner:2021,Hudson:2012}. 

Recent observational progress has significantly increased the relevance of velocity-based analyses. In particular, the Peculiar Velocity Survey of the Dark Energy Spectrograph Instrument (DESI) has delivered the most precise measurements of galaxy peculiar velocities and the tightest constraints on $\fs$ at $z\sim 0$ \cite{Lai:2025a, Turner:2025b, Qin:2025c, Carr:2025d}. In parallel, peculiar velocities leave distinct imprints on the cosmic microwave background via the kinetic Sunyaev–Zel’dovich (kSZ) effect \cite{Sunyaev:1980,Ostriker:1986}, providing an independent avenue to probe large-scale structure and gravity \cite{Hernandez:2006,Okumura:2014,Mueller:2015,Sugiyama:2016,Sugiyama:2017,Zheng:2020,Hand:2012,Bernardis:2017,Aghanim:2018,Sugiyama:2018,Li:2018,Calafut:2021,Chaves:2020,Chen:2021}. These developments place renewed demands on accurate theoretical modeling of velocity statistics on large scales \cite{Peebles:1980,Fisher:1995,Burkey:2004,Bhattacharya:2008,Okumura:2014,Koda:2014,Howlett:2019,Tonegawa:2024,Chen:2025}.

A notable feature of the velocity field is its enhanced sensitivity to long-wavelength perturbations. In linear theory, velocities scale as $\bfv(\bfk)\propto (i\bfk/k^2)\delta(\bfk)$ in Fourier space, amplifying the contribution of large-scale modes relative to density fluctuations. As a result, velocity correlation functions retain significant signal on large scales even beyond baryon acoustic oscillation scales, $r> 100\mpcoh$ \cite{Okumura:2019}. However, this same sensitivity renders velocity statistics particularly vulnerable to systematic effects associated with finite observational or simulation volumes.

Indeed, several studies have reported that velocity correlation functions measured from cosmological $N$-body simulations exhibit systematically suppressed amplitudes on large scales compared to theoretical predictions \cite{Sheth:2009,Okumura:2014,Okumura:2019}. This discrepancy persists for dark matter halos and galaxies, which are more directly relevant to observations. Despite its clear presence, the origin and cosmological impact of this suppression have not been fully quantified, especially in the context of simulation-based modeling and parameter inference.  

Simulation-based approaches, such as emulators built from suites of cosmological $N$-body simulations, have become indispensable tools for precision cosmology \cite{Heitmann:2009,Heitmann:2013,Okumura:2015,Lawrence:2017,Euclid:2019,Nishimichi:2019,Kobayashi:2020,Koyama:2025}. While highly successful for density-based statistics, these methods are inherently limited by the finite size of simulation boxes. Periodic boundary conditions eliminate modes with wavelengths larger than the box size, leading to missing long-wavelength power \cite{Hu:2003,Putter:2012}. Such super-survey effects are known to alter large-scale correlations and covariance matrices \cite{Bagla:2005,Hamilton:2006,Takada:2013,Carron:2014,Li:2014a,Li:2014b}, and are expected to be especially severe for velocity statistics due to their strong infrared sensitivity. Unlike real galaxy surveys, whose window functions can be explicitly modeled and deconvolved, cosmological simulations impose a hard cutoff on Fourier modes larger than the box size. This cutoff is imprinted in the initial conditions and persists throughout the nonlinear evolution. As a result, simulation-based predictions inherently lack contributions from long-wavelength modes that are present in the real Universe.

In this Letter, we investigate how the finite volume of cosmological simulations systematically biases measurements of halo velocity correlation functions on large scales. Using $N$-body simulations with different box sizes, we demonstrate that the suppression arises from missing long-wavelength modes and depends on the total simulation volume rather than the size of the analyzed subregion. We show that introducing an effective large-scale cutoff in the theoretical modeling recovers unbiased constraints on $\fs$, while neglecting this effect can lead to parameter biases. Our results highlight a previously underappreciated limitation of simulation-based velocity modeling and have direct implications for future analyses of peculiar velocity and kSZ data.

Throughout this Letter, we assume the spatially flat $\Lambda$CDM model as our fiducial model \cite{Planck-Collaboration:2016}: $\Omega_m = 1 - \Omega_\Lambda = 0.315 $, $\Omega_b=0.0492$, $H_0 = 67.3~[{\rm km/s/Mpc}]$, $n_s=0.965$, and the present-day value of $\sigma_8(z)$ to be $\sigma_{8}(0)=0.8309$. 
\\ \\
{\it Velocity Autocorrelation Function.}
We consider the autocorrelation function of the peculiar velocity field $\bfv(\bfx)$ \cite{Peebles:1980,Gorski:1989}, which is most severely affected by the finite volume effect. Since the peculiar velocities are sampled at galaxy positions, they are a density-weighted quantity, often referred to as the momentum density field \cite{Okumura:2014}, $[1+\delta_{h}(\bfx)] \bfv(\bfx)$, where $\delta_{h}(\bfx)$ is the overdensity field. One can observe only the line-of-sight components of the velocity field, $v_{\parallel}(\bfx) = \bfv(\bfx)\cdot \hat{\bfx}$, where a hat denotes a unit vector, i.e., $\hat{\bfx}=\bfx/|\bfx|$. Thus, the velocity autocorrelation function has an angular dependence and is defined by 
\begin{equation}
\label{eq:vv}
\xivv(\bfr) = \langle[1+\delta_{h}(\bfx_{1})][1+\delta_{h}(\bfx_{2})] v_{\parallel}(\bfx_1) v_{\parallel}(\bfx_2)\rangle \, ,
\end{equation}
where $\bfr = \bfx_{2}-\bfx_{1}$. 

Since we are interested in large scales, we rely on linear theory throughout this Letter. The velocity field is related to the density field through the continuity equation \cite{Fisher:1995,Strauss:1995}. The autocorrelation function of the line-of-sight velocity field has an angular dependence of $\mu^0$ and $\mu^2$ in linear theory, where $\mu$ is the cosine of the angle between the line-of-sight and separation vector, $\mu = \hat{\bfx}\cdot\hat{\bfr}$, under the distant-observer approximation. Thereby, $\xivv$ has nonzero monopole and quadrupole moments. Throughout this Letter, we mainly investigate the angularly-averaged monopole moment, denoted by $\xivv(r)$, because it is the most significantly affected by the finite-volume effect. Investigating the quadrupole moment is beyond the scope of this Letter. We present a joint analysis of the monopole and quadrupole moments in future work.

The velocity autocorrelation function is expressed by the Hankel transform of the matter power spectrum $\Pmm(k)$,
\begin{align}
\xivv(r) &=
\frac{a^2H^2f^2}{3} \int^{\infty}_{0} \frac{dk}{2\pi^2} \Pmm(k)   j_{0}(kr) W(k) , \label{eq:xiv0}
\end{align}
where $a$ is the scale factor, $H$ is the Hubble parameter, $f$ is the linear growth rate, $j_{\ell}$ are the $\ell$th-order spherical Bessel functions. Generally, the window function, $W(k)$, is chosen to avoid $k=0$ in the integral, namely, $W(k)=\Theta(k)$, where $\Theta$ is the step function. In this work, we examine how the finite size of simulation boxes affects the measured velocity correlation function. Numerical simulations carried out in a finite cubic box with periodic boundary conditions can only represent discrete Fourier modes, corresponding to integer multiples of the fundamental mode $2\pi/L$. In this setup, the Cartesian directions (i.e., $x$, $y$, and $z$ directions) are special, and the resulting cutoff is inherently anisotropic. For simplicity, however, in the analytical model, we introduce an isotropic cutoff parameter, $\kmin$, and modify the window function as 
\be
W(k) = \Theta(k-\kmin) \, .
\ee
Eq.~(\ref{eq:xiv0}) then becomes 
\begin{align}
\xivv(r) &=
\frac{a^2H^2f^2}{3} \int^{\infty}_{\kmin} \frac{dk}{2\pi^2} \Pmm(k)j_{0}(kr). \label{eq:xiv0kmin}
\end{align}
We use the \texttt{CAMB} package to compute $\Pmm(k)$ \citep{Lewis:2000}. While we adopt the nonlinear prediction based on the revised \texttt{HALOFIT} \citep{Smith:2003,Takahashi:2012}, the nonlinerities at the scales we consider are negligible as we see later. Since $\Pmm(k)$ is proportional to the square of the normalization parameter of the density fluctuation, $\sigma^2_8(z)$, the velocity correlation function depends on the combination, $(\fs)^2$. Accordingly, we treat $\fs$ as a single free parameter, and simultaneously allow $\kmin$ to vary when we fit the model to simulation data.

The velocity field in redshift space is equivalent to that in real space for the linear theory limit \cite[e.g.,][]{Okumura:2014}. Since we are interested in the finite volume effect on large scales, we do not consider the RSD effect throughout this Letter.
\\ \\
{\it N-body Simulations.}
To investigate the impact of finite simulation volumes on the velocity statistics, we use two sets of $N$-body simulations with different box sizes, run as a part of the \darkquest project \citep{Nishimichi:2019}. Each simulation contains $n_p=2048^3$ particles in cubic boxes with side lengths of $1\,\Gpcoh$ and $2\,\Gpcoh$, hereafter referred to as the $\LoneG$ and $\LtwoG$ simulations, respectively. We analyze $N_{\rm real} =24$ and $8$ independent realizations for the $\LoneG$ and $\LtwoG$ simulations, respectively, at redshift $z=0.306$. Dark matter halos and subhalos are identified with the \rockstar algorithm \cite{Behroozi:2012,Behroozi:2013}. The velocity of the subhalo is determined by the average particle velocity within the innermost 10 per cent of the subhalo radius. To mimic a realistic galaxy distribution, we populate (sub)halos with galaxies according to a halo occupation distribution model determined for a LOWZ galaxy sample of the SDSS-III Baryon Oscillation Spectroscopic Survey \cite{Parejko:2013}. Throughout this work, we treat the subhalo-based galaxy sample as a practical proxy for realistic velocity tracers. See Table III of Ref.~\cite{Okumura:2024} for the details of the mock galaxy sample. 
 
The correlation function of the line-of-sight velocity of halos is estimated as
\be
\xivv(r) = \frac{\sum_{i,j}v_\parallel(\bfx_i)v_\parallel(\bfx_j)}{RR(\bfr)}, \label{eq:estimator}
\ee
where the sum is taken over all pairs, weighted by the product of the line-of-sight velocities. The denominator, $RR(r)$, is the pair count of the random distributions as a function of the separation $r$, which can be evaluated analytically owing to the simple cubic geometry of the simulation volume. We regard each direction along the three axes of the simulation boxes as the line of sight, and the statistics are averaged over the three projections of all $N_{\rm real}$ realizations.

We need to construct a covariance matrix for the measured velocity correlation function for a statistical analysis. Since the number of independent realizations is not sufficient, particularly for the $\LtwoG$ simulation, we adopt a bootstrap resampling method \cite{Barrow:1984}. We adopt the number of resampling for each realization to be $N_{\bs}=20$ and 30 for $\LoneG$ and $\LtwoG$ simulations, respectively. Based on the standard decomposition of the total covariance, we separate the total covariance matrix into the covariance from independent realizations, $\Cext$, and that from the bootstrap resampling, $\Cint$, \cite{Norberg:2009}
\begin{equation}
    \mathbf{C} = \Cext+\Cint,  
\end{equation}
where $\Cext$ and $\Cint$ quantify the fluctuations among independent realizations (sample variance) and the average covariance due to sampling noise and small-scale fluctuations within individual realizations, respectively. The former is estimated as 
\be
\left(\Cext\right)_{ij}
= \frac{1}{N_{\real}-1}\sum_{k=1}^{N_{\real}}
\Delta^k_i\Delta^k_j \, ,
\ee
where $\Delta^k_i = \xivv^k(r_i)-\bar{\xivv}(r_i)$, $\xivv^k(r_i)$ is the velocity correlation function measured in the $k$-th realization at separation $r_i$, and $\bar{\xivv}$ denotes the mean of the measurements across the independent realizations. The latter is given by
\be
    \left(\Cint\right)_{ij}
    = \frac{1}{N_{\real}}\sum_{k=1}^{N_{\real}}
    \frac{1}{N_{\bs}-1} 
    \sum_{b=1}^{N_{\bs}} 
    \Delta^{k,b}_i\Delta^{k,b}_j \, ,
\ee
where $\Delta^{k,b}_i=\xivv^{\,k,b}(r_i)-\bar{\xivv}^{k}(r_i)$, $\xivv^{\,k,b}(r_i)$ is the velocity correlation function measured in the $b$-th bootstrap resample of the $k$-th realization, and $\bar{\xivv}^{k}$ denotes the bootstrap mean within box $k$. Since we use the mean of the correlation functions measured across independent realizations, our total covariance is divided by the number of independent realizations to represent the error of the mean, $\mathbf{C}\rightarrow \mathbf{C}/N_{\rm real}$. The error bars for the measured correlation function are the square root of the diagonal parts of the covariance matrix. 

\begin{table}[tb]
    \centering
    \caption{
    Properties of our mock halo catalogs. The quantity $m_p$ is the particle mass, $N_{\rm real}$ and $N_{\rm bs}$ are the numbers of realizations and bootstrap resampling, respectively. 
    }
    \begin{tabular}{lcccccc}
    \hline\hline
    & \# of & $m_p$ & (Sub)box size & 
    & &
     \\ 
     & runs&$[h^{-1}M_{\odot}]$& $[\Gpcoh]$ & $N_{\rm real}$&& $N_{\rm bs}$ \\
     \hline
    $\LoneG$       & 24  & $1.02 \times 10^{10} $ & 1   & 24  &&  20 \\
     \hline
    $\LtwoG$       &  8  & $8.16 \times 10^{10} $ & 2   &  8  &&  30 \\
    $\LoneGsub$    &     &                        & 1   &  64  &&  30 \\
    $\Lthsub$      &     &                        & 0.3 &  1728  &&  30 \\
    \hline\hline
    \end{tabular}
    \label{tab:realizations}
\end{table}

Besides the standard analysis above, to further investigate the effect of long-wavelength modes beyond the simulation box, we subdivide each $\LtwoG$ simulation into smaller sub-boxes with side lengths of half of the parent box, labeled as the $\LoneGsub$ simulation. The box size is chosen to coincide with the $\LoneG$ simulation. The number of the realizations for the subvolume is $N_{\rm real} =8\times 2^3 = 64$. Additionally, we create even smaller subvolumes with $15\%$ of the side length, denoted by $\Lthsub$, with the total number of realizations of $N_{\rm real} =8\times 6^3 = 1728$. Since the periodic boundary condition does not hold for the analysis of the subvolumes, the pair count $RR$ in Eq.~(\ref{eq:estimator}) no longer admits a simple analytic expression and is therefore evaluated numerically using random points.\footnote{Using the original $\LoneG$ and $\LtwoG$ simulations, we have confirmed that for estimating $RR$, the analytical calculation and direct counting of random points give consistent results.} 

\begin{figure}[t]
\begin{center}
\includegraphics[width = 0.4\textwidth]{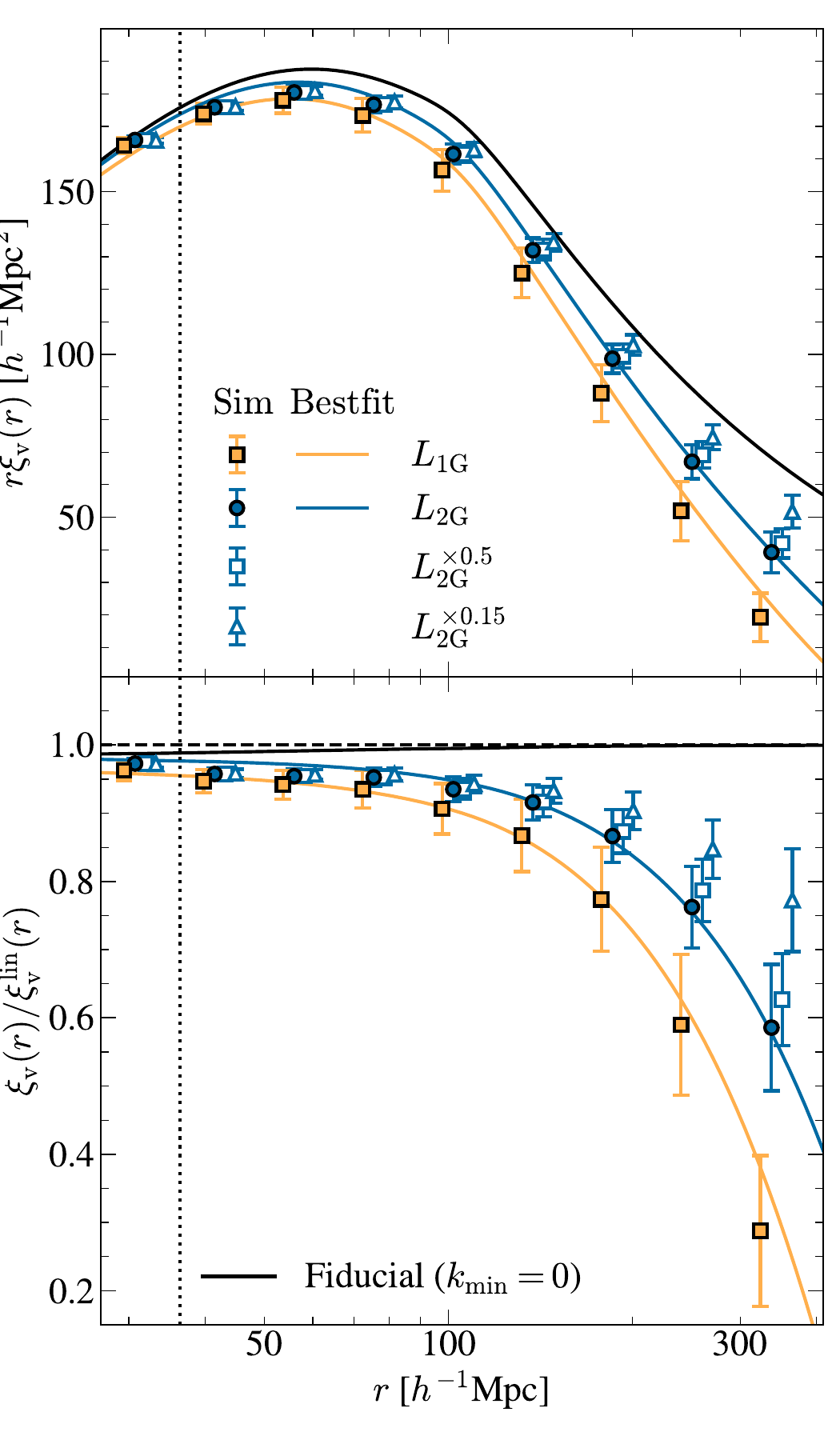}
\caption{Upper panel: angularly-averaged velocity correlation functions of halos, $\xivv$. The filled orange squares and blue circles are the measurements from the cubic boxes with side length of $1\Gpcoh$ ($\LoneG$) and $2\Gpcoh$ ($\LtwoG$), respectively. The open squares and triangles are the results from the subdivided simulation boxes, $\LtwoG$, to the side lengths of $1\Gpcoh$ ($\LoneGsub$) and $0.3\Gpcoh$ ($\Lthsub$), respectively. The black curves are the nonlinear model with the true parameter of $\fs$ with $\kmin\to 0$. The orange and blue curves are the best-fitting model with $(\fs,\kmin)$ at $36\leq r\leq 400\mpcoh$ for $\LoneG$ and $\LtwoG$ simulations, respectively (see Table \ref{tab:table}).
Lower panel: ratios of the above results to the linear-theory prediction with the true parameter of $\fs$ with $\kmin\to 0$.
}
\label{fig:vv0_measurements}
\end{center}
\end{figure}

The details of the original periodic $\LoneG$ and $\LtwoG$ realizations, as well as the subvolumes of $\LtwoG$ realizations, are summarized in Table \ref{tab:realizations}.
\\ \\
{\it Velocity Correlation Measurements.} The upper panel of Fig. \ref{fig:vv0_measurements} shows the angularly-averaged velocity correlation function, $\xivv(r)$. Clearly, the amplitude of the correlation functions measured from both the $\LtwoG$ and $\LoneG$ is systematically lower than the theoretical prediction with the fiducial model, where $\kmin= 0$. To highlight the difference, we plot these results divided by the linear theory prediction with the fiducial model. The amplitude of the measurements is lower than the linear theory prediction by $\sim5-10\%$ at around $100\mpcoh$, with the discrepancy increasing at larger scales. The scale dependence of the suppression closely follows expectations from missing long-wavelength power, with deviations becoming increasingly pronounced at separations approaching a substantial fraction of the simulation box size. This trend is consistent with earlier findings \citep{Sheth:2009,Okumura:2014}. The suppression is found more pronounced in the smaller $\LoneG$ volume, reflecting the dependence on the minimum accessible wavenumber, $\kmin$.

In contrast, when we use the $\LoneGsub$ realizations, which were subdivided from the $\LtwoG$ boxes to have the same volumes as the $\LoneG$ boxes, the suppression becomes similar to the $\LtwoG$ case. It is because the effect of the Fourier modes longer than the sub-boxes remains even after subdividing the original simulation boxes \cite{Takada:2013,Li:2014a,Chiang:2014}. For an additional check, even from the smaller sub-boxes, $\Lthsub$, we observe the suppression consistent with the original $\LtwoG$-box case.  
\\ \\
{\it Cosmological Analysis.}
Given the significant discrepancies between the measured velocity correlations and theoretical predictions on large scales, we assess how accurately the model constrains the cosmological parameter, $\fs$, by varying the minimum wavenumber $\kmin$. We use the smaller, $1\Gpcoh$ simulations, with which the measurement of the velocity correlation exhibits larger deviation from linear theory. 

\begin{figure}[tbp]
\begin{center}
\includegraphics[width = 0.4\textwidth]{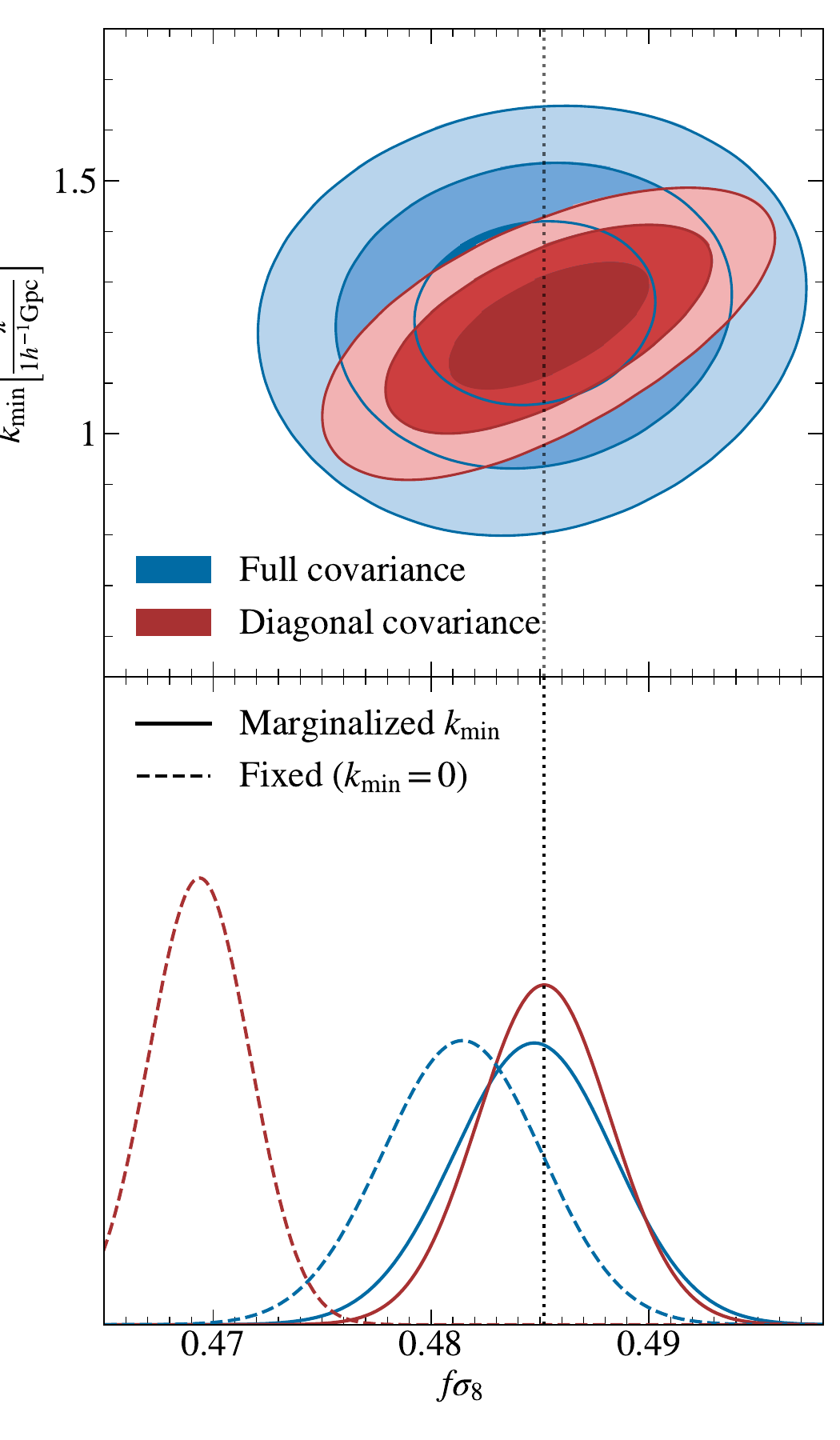}
\caption{ 
Top: joint constraints on the growth rate $\fs$ and the minimum wavenumber $\kmin$ in the $\LoneG$ simulation box. The contours show the $1\sigma$, $2\sigma$, and $3\sigma$ confidence levels from inward. The blue and red contours correspond to the results using the full and diagonal covariance matrices, respectively. 
Bottom: One-dimensional posterior distributions for $\fs$. The solid curves show the result when $\kmin$ is marginalized over, while the dashed curves show that where $\kmin$ is fixed to $0.1\pi/\LoneG\simeq 3\times 10^{-4} [\hompc]$. The vertical dotted line indicates the fiducial input value of $\fs$.
}
\label{fig:vv0_contours}
\end{center}
\end{figure}
%
\begin{figure}[t]
\begin{center}
\includegraphics[width = 0.4\textwidth]{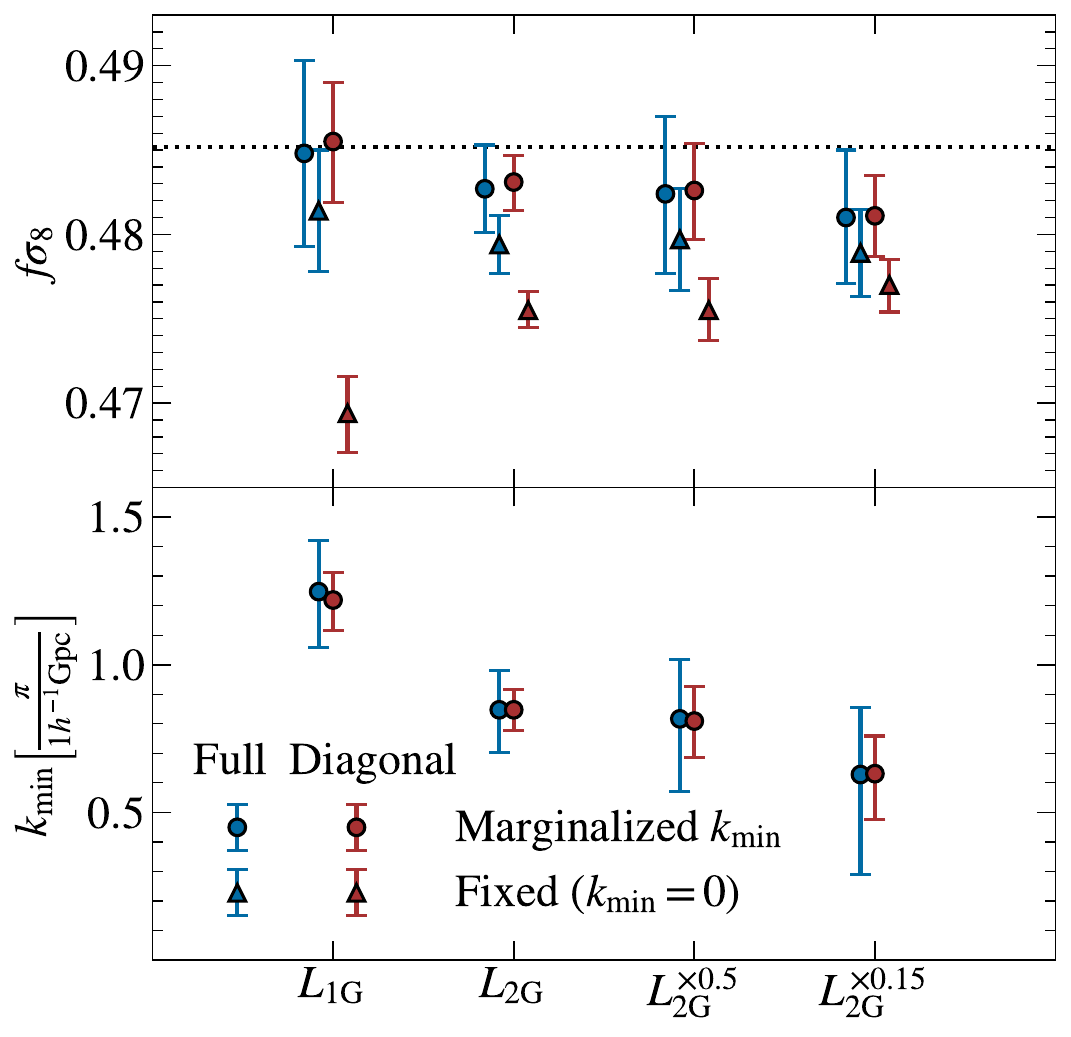}
\caption{
Constraints on model parameters from simulation boxes with different sizes and subdivided boxes. The upper and lower panels show the constraints on the growth rate $\fs$ and the scale-cut parameter $\kmin$. The blue and red points show results obtained using the full and diagonal covariance matrices, respectively. In the upper panel, the triangles show the results when we fix $\kmin=0$. The true value of $\fs$ from the simulations is indicated by the horizontal dotted line. 
}
\label{fig:best_fitting_diff_box_vv0}
\end{center}
\end{figure}

The $\chi^2$ statistic is given by
\begin{align}
\chi^{2}(\fs,\kmin) = \sum\limits_{i,j}^{N_{\rm bin}} & 
\Delta_{i}
C^{-1}_{ij} 
\Delta_{j}, \label{eq:chi2}
\end{align}
where $\Delta_{i}= \xivv(r_{i}) - \xivv^{\rm th}(r_{i};\fs,\kmin)$ is the difference between the measured correlation function and the theoretical prediction with $(\fs,\kmin)$ being a parameter set to be constrained, $C^{-1}$ denotes the inverse of the covariance matrix for the mean of the correlation function, and $N_{\rm bin}$ is the number of bins of the velocity correlation monopole. We use the data only on large scales, since the scales where nonlinear effects dominate are not of our interest. We thus choose the range of $36\leq r\leq 400\mpcoh$ and $N_{\rm bin}=8$.

The upper panel of Fig.~\ref{fig:vv0_contours} shows the joint constraints on $(\fs,\kmin)$ from the $\LoneG$ simulations. For comparison, both results, using all elements and only the diagonals of the covariance matrix, are shown. These two results are in very good agreement, although the latter constraint is much tighter, as expected. The lower panel of Fig. \ref{fig:vv0_contours} shows the posterior probability distributions of $\fs$ with $k_{\min}$ marginalized over. For both cases, the best-fitting value of $\fs$ is fully consistent with the true cosmological model assumed for running the simulations, $\fs = 0.4852$.

In the bottom panel, we also show the $\fs$ constraints when $\kmin$ is fixed close to the $\kmin \to 0$ limit by choosing $\kmin=0.1\pi / \LoneG \simeq 3\times 10^{-4}\hompc$. In this case, using only the diagonal elements of the covariance matrix results in a significantly lower value of $\fs$. However, once the full covariance matrix is utilized, the shift of the best-fitting value of $\fs$ still occurs, but the constraint remains within the $1\sigma$ confidence level. The one-dimensional posteriors are also shown on the left-hand side of Fig.~\ref{fig:best_fitting_diff_box_vv0} and the first row of Table~\ref{tab:table}.

These results indicate that when considering simulation-based modeling of velocity correlation functions, the effect of the finite volume of the simulation boxes must be carefully taken into account. If the lower limit of the integral in the Hankel transform, $\kmin$, is treated as a free parameter, the simulation-based approach provides an accurate prediction of the velocity statistics. Even if we ignore the finite volume effect by setting $\kmin=0$, as long as the full covariance matrix is used for the analysis, the parameter $\fs$ can be unbiasedly recovered, at least at the precision investigated in this Letter. 

\begin{table*}[hptb]
\begin{center}
\caption{Summary of constraints on $\fs$. The results are shown for both cases when using the diagonal elements of the covariance matrix and the full covariance. The degrees of freedom are $\nu=6$ and $\nu=7$ when $\kmin$ is varied and fixed to $\kmin=0$, respectively.}
\begin{tabular} {l c c c c c c c c c c c c c c} 
    \hline\hline
    & \multicolumn{6}{c}{Diagonal-only} && \multicolumn{6}{c}{Full covariance}\\ 
    \cline{2-7} 
    \cline{9-14}
    & \multicolumn{3}{c}{Varied $\kmin$} && \multicolumn{2}{c}{Fixed $(\kmin =0)$} && \multicolumn{3}{c}{Varied $\kmin$} && \multicolumn{2}{c}{Fixed $(\kmin=0)$}\\
    \cline{2-4}
    \cline{6-7}
    \cline{9-11}
    \cline{13-14}
    & $\fs/\fs^{\rm fid}$ &  $k_{\min}\LoneG/\pi$  &  $\chi^{2}/\nu$  && $\fs/\fs^{\rm fid}$ &  $\chi^{2}/\nu$ && $\fs/\fs^{\rm fid}$ &  $k_{\min}\LoneG/\pi$  &  $\chi^{2}/\nu$  && $\fs/\fs^{\rm fid}$ &  $\chi^{2}/\nu$ \\ 
    \hline
    $\LoneG$    & $1.0006_{-0.0095}^{+0.0093}$ & $1.22_{-0.13}^{+0.12}$  & 0.076 && $0.9674_{-0.0062}^{+0.0060}$ & 6.158 && $0.9990_{-0.015}^{+0.014}$ & $1.25_za{-0.24}^{+0.22}$  & 0.076 && $0.9922_{-0.0095}^{+0.0095}$ & 3.652 \\ 
 
    $\LtwoG$    & $0.9957_{-0.0035}^{+0.0033}$ & $0.847_{-0.071}^{+0.069}$ & 0.117 && $0.9800_{-0.0021}^{+0.0023}$ & 6.519 && $0.9948_{-0.0054}^{+0.0054}$ & $0.85_{-0.15}^{+0.13}$ & 0.125 && $0.9880_{-0.0035}^{+0.0035}$ & 3.131 \\ 
  
    $\LoneGsub$ & $0.9946_{-0.0060}^{+0.0058}$ & $0.81_{-0.12}^{+0.12}$  & 0.021 && $0.9800_{-0.0018}^{+0.0019}$ & 2.298 && $0.9942_{-0.0097}^{+0.0095}$ & $0.82_{-0.25}^{+0.20}$  & 0.021 && $0.9887_{-0.0062}^{+0.0062}$ & 1.253 \\ 
    
    $\Lthsub$   & $0.9915_{-0.0049}^{+0.0049}$ & $0.63_{-0.15}^{+0.13}$  & 0.012 && $0.9831_{-0.0033}^{+0.0031}$ & 0.852 && $0.9913_{-0.0080}^{+0.0082}$ & $0.63_{-0.34}^{+0.23}$  & 0.014 && $0.9870_{-0.0054}^{+0.0054}$ & 0.434 \\

    \hline\hline
 
    \label{tab:table}
\end{tabular}
\end{center}
\end{table*}

In Fig.~\ref{fig:best_fitting_diff_box_vv0} and Table~\ref{tab:table}, we also show the constraints on $\fs$ and $\kmin$ obtained from the larger-box simulations, $\LtwoG$. We find a trend similar to the case of the smaller, $\LoneG$ simulations, with the best-fitting value of $\kmin$ being smaller, about a half, as expected. Even though we use the subdivided boxes, $\LoneGsub$ and $\Lthsub$, the trend is found to persist. Although the box size of $\LoneG$ and $\LoneGsub$ is the same, the best-fitting values of $\kmin$ are different because the Fourier modes beyond $\LoneG=1\Gpcoh$ are imprinted in the $\LoneGsub$ subboxes. These results demonstrate that biases in $\fs$ can arise not from deficiencies in the velocity data themselves, but from an incomplete treatment of large-scale modes in the theoretical modeling.
%
\begin{figure}[t]
\begin{center}
\includegraphics[width = 0.4\textwidth]{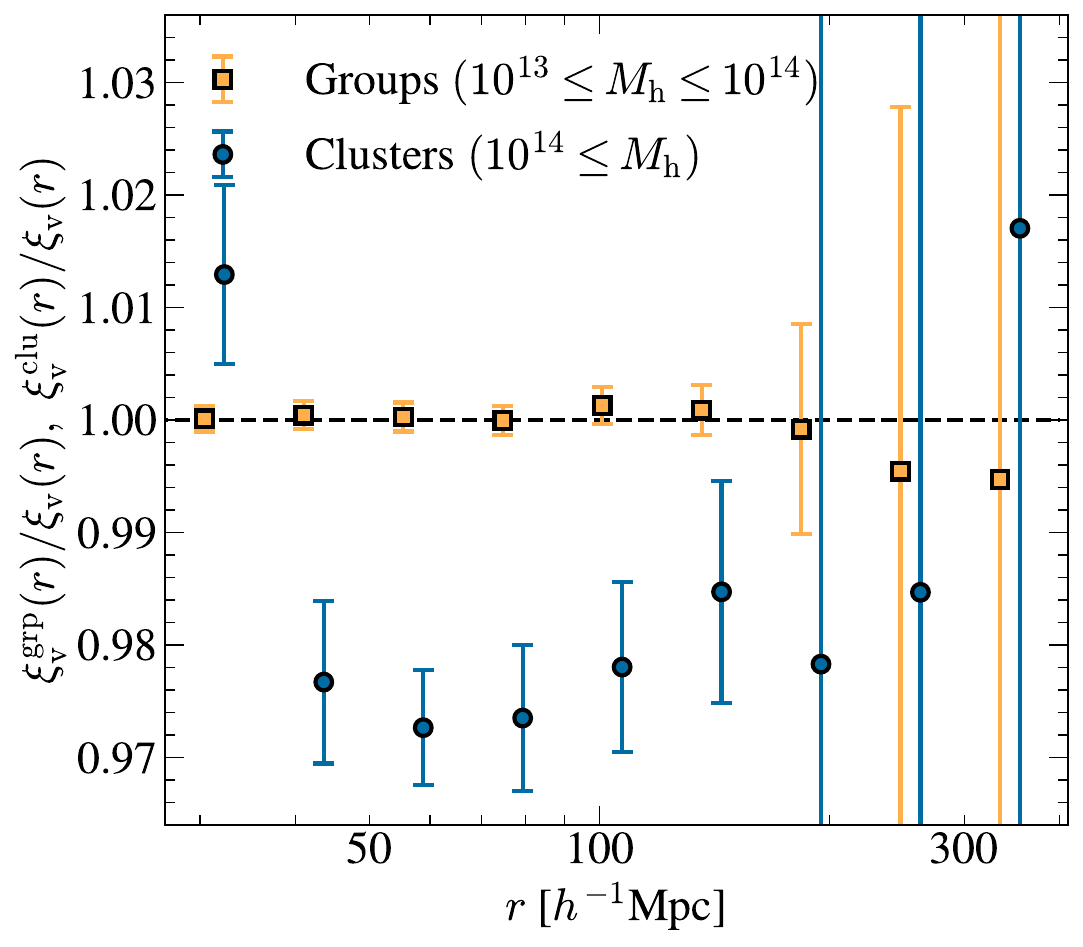}
\caption{
Ratios of the velocity correlation functions of halos with certain mass ranges and those of subhalos determined from the HOD modeling of the SDSS-III BOSS LOWZ galaxy sample. The host subhalo mass of the LOWZ sample is very similar to that of the group subsample. 
}
\label{fig:mass}
\end{center}
\end{figure}
\\ \\
{\it Halo-mass Dependence.}
Via the continuity equation, the cosmological velocity field directly traces the matter density perturbations. If the amplitude of the velocity correlation function of biased tracers is different from that of the underlying dark matter, such a difference is called the velocity bias \cite{Carlberg:1990,Colin:2000,Yoshikawa:2003,Baldauf:2015,Zheng:2015}. Ref.~\cite{Okumura:2014} found that the velocity correlation of dark matter is indeed more severely affected by the finite-volume effect than that of halos. Furthermore, if one finds a discrepancy in the velocity correlation amplitudes between two different galaxy samples, it can be evidence for the violation of Einstein's equivalence principle. Thus, when the simulation-based modeling is adopted to test it, the simulation-volume effect needs to be carefully interpreted. 

To see the effect, we use the host halo sample and split it into different mass subsamples, $M_h\leq 10^{13}$ and $10^{13}\leq M_h\leq 10^{14}$. They correspond to the halos that typically represent galaxy groups and clusters, respectively. These samples were used and summarized in Ref.~\cite{Okumura:2024}. We measure the velocity autocorrelation functions from these subsamples, $\xivv^{\rm grp}$ and $\xivv^{\rm clu}$, and show the ratios to the autocorrelation function of our main subhalo samples in Fig.~\ref{fig:mass}. The average mass of the subhalo sample is $2.5\times10^{13}\moh $, very similar to that of the group subsample, exhibiting the ratio of their correlation functions equal to unity. On the other hand, the result from the cluster-size massive halos shows systematically lower amplitude. The trend is consistent with the earlier work of Ref.~\cite{Okumura:2014}. When the galaxies in smaller-mass halos are analyzed, the difference in the correlation amplitude from the cluster-sized halos would become more significant \cite{Okumura:2014}. 

The observed mass dependence implies that finite-volume effects and velocity bias cannot be cleanly separated when analyzing velocity correlations of biased tracers. In particular, comparisons of velocity statistics across different halo or galaxy samples must account for the fact that finite simulation volumes can modulate the apparent amplitude differences. This is especially relevant for tests of the Einstein equivalence principle using simulation-calibrated velocity models, where an unaccounted finite-volume effect could mimic or obscure genuine physical signals.
\\ \\
{\it Conclusions.} 
In this work, we have investigated the impact of finite simulation volumes on measurements and cosmological inference based on the halo peculiar velocity correlation function on large scales. Using suites of $N$-body simulations with different box sizes, together with controlled sub-volume analyses, we demonstrated that the absence of long-wavelength modes in finite simulation boxes leads to a systematic suppression of the velocity correlation function relative to linear-theory predictions. This effect becomes increasingly pronounced as the total simulation volume decreases.

By explicitly comparing independent simulations and sub-volumes drawn from larger boxes, we showed that the suppression is governed by the availability of long-wavelength modes rather than by the physical size of the region used for the measurement. Sub-volumes carved out of larger simulations retain the imprint of modes longer than their own extent, resulting in velocity correlations that differ qualitatively from those measured in truly independent small boxes. This behavior highlights a fundamental distinction between finite-volume effects in simulations and those arising from limited survey geometries in observational data.

We quantified how these finite-volume effects propagate into cosmological parameter inference. Likelihood analyses of the velocity correlation reveal that fixing the lower limit of the Hankel transform to approximate the $\kmin\to0$ limit can lead to biased estimates of the growth rate parameter $\fs$, particularly when correlations between separation bins are neglected. Allowing the effective large-scale cutoff scale $\kmin$ to vary absorbs the impact of missing long-wavelength modes and enables unbiased recovery of $\fs$ at the precision explored in this work. We emphasize that treating $\kmin$ as a free parameter should be viewed as a diagnostic tool rather than a complete physical model, as it encapsulates finite-volume systematics inherent to simulation-based predictions.

These findings have direct implications for simulation-based analyses of velocity statistics, including emulator-driven approaches. While real galaxy surveys are not subject to a hard box scale, simulations used as theoretical templates inevitably are. Finite-volume imprints in such simulations can therefore bias cosmological constraints if not properly modeled or marginalized over. Our results highlight the importance of consistently accounting for missing long-wavelength modes when comparing simulations with theoretical predictions. While the underlying dynamics in the real Universe are not subject to such a cutoff, using finite-volume simulations as direct theoretical templates for observations, without accounting for the missing long-wavelength modes, can lead to biased interpretations of peculiar velocity or kSZ measurements.

Several extensions of this work merit further investigation. A natural next step is to study finite-volume effects on the quadrupole moment of the velocity correlation function, which requires introducing anisotropic large-scale cutoffs rather than the isotropic treatment adopted here. Incorporating nonlinear effects and performing the analysis in redshift space will also enable a more realistic modeling of velocity statistics, including Finger-of-God contributions. In addition, the velocity–density cross-correlation offers a complementary observable that may help mitigate finite-volume systematics while preserving sensitivity to large-scale modes. Together, these developments will help ensure that velocity-based cosmological analyses fully exploit the growing precision of current and upcoming observational data while maintaining robust control over simulation-induced systematics.
\\ \\
{\it Acknowledgments.}
We thank Ryuichiro Hada and Masahiro Takada for useful conversations. 
T.~O. acknowledges support from the Taiwan National Science and Technology Council under Grants 
Nos. NSTC 112-2112-M-001-034-, 
NSTC 113-2112-M-001-011- and
NSTC 114-2112-M-001-004-, and the Academia Sinica Investigator Project Grant No. AS-IV-114-M03 for the period of 2025–2029. T.N~. is supported by MEXT/JSPS KAKENHI Grant Numbers JP22K03634, JP24H00215, and JP24H00221. Numerical simulations were carried out on Cray XC50 at the Center for Computational Astrophysics, National Astronomical Observatory of Japan.
\\ \\
{\it Data availability.}
The data are not publicly available. The
data are available from the authors upon reasonable request.

\bibliography{main_prd.bbl}

\end{document}